\documentclass[prc,amsmath,twocolumn,showkeys,showpacs,superscriptaddress]{revtex4}

\usepackage{gensymb}
\usepackage{graphicx,color}
\usepackage{amssymb}
\usepackage{enumerate}
\usepackage{verbatim}
\usepackage{natbib}
\usepackage{amsmath}

\begin{document}
  \newcommand {\nc} {\newcommand}
  \nc {\Sec} [1] {Sec.~\ref{#1}}
  \nc {\IR} [1] {\textcolor{red}{#1}} 
  \nc {\IB} [1] {\textcolor{blue}{#1}}

\title{Three-body model for the two-neutron decay of $^{16}$Be}

\author{A.~E.~Lovell}
\affiliation{National Superconducting Cyclotron Laboratory, Michigan State University, East Lansing, MI 48824, USA}
\affiliation{Department of Physics and Astronomy, Michigan State University, East Lansing, MI 48824, USA}
\author{F.~M.~Nunes}
\affiliation{National Superconducting Cyclotron Laboratory, Michigan State University, East Lansing, MI 48824, USA}
\affiliation{Department of Physics and Astronomy, Michigan State University, East Lansing, MI 48824, USA}
\author{I.~J.~Thompson}
\affiliation{Lawrence Livermore National Laboratory, L-414, Livermore, CA 94551, USA}

\date{\today}


\begin{abstract}
\begin{description}

\item[Background:]  While diproton decay was first theorized in 1960 and first measured in 2002, it was first observed only in 2012. The measurement of $^{14}$Be in coincidence with two neutrons suggests that  $^{16}$Be does decay through the simultaneous emission of two strongly correlated neutrons.  
\item[Purpose:]  In this work, we construct a full three-body model of $^{16}$Be (as $^{14}$Be + n + n) in order to investigate its configuration in the continuum and in particular the structure of its ground state.  
\item[Method:]  In order to describe the three-body system, effective n-$^{14}$Be potentials were constructed, constrained by the experimental information on $^{15}$Be. The  hyperspherical R-matrix method was used to solve the three-body scattering problem, and the resonance energy of $^{16}$Be was extracted from a phase shift analysis.    
\item[Results:]  In order to reproduce the experimental resonance energy of $^{16}$Be within this three-body model, a three-body interaction was needed.  For extracting the width of the ground state of $^{16}$Be, we use the full width at half maximum of the derivative of the three-body phase shifts and the width of the three-body elastic scattering cross section.  
\item[Conclusions:] Our results confirm a dineutron structure for $^{16}$Be, dependent on the internal structure of the subsystem $^{15}$Be.  
\end{description}
\end{abstract}



\maketitle
\section{Introduction}


Exotic nuclei are found across the nuclear chart.  Proton and neutron halos are found near the proton and neutron dripline, respectively, not only in the lightest mass nuclei but also possibly in nuclei as heavy as neon \cite{Tanihata2013}.  Two-nucleon halo systems can be Borromean, where, if we think of these in terms of a core plus two neutrons or protons, the three-body system is bound but each of the two subsystems is unbound \cite{ReactionsBook} (Ch. 9).  Unsurprisingly, beyond the dripline, novel structures can give rise to exotic decay paths.  

Two-proton decay was first theorized in 1960 \cite{GOLDANSKY1960482}.  When two nucleons decay from a core, there are three possible mechanisms.  First, the two nucleons can decay simultaneously, in a true three-body decay.  If there is a state in the A-1 nucleus below the ground state of the parent nucleus, the two are likely to decay sequentially, stepping through the intermediate A-1 state.  However, if the ground state in the A-1 nucleus is energetically inaccessible to the decay of  two nucleons and there is correlation between the two nucleons before the decay, dinucleon decay is the likely alternative.  

Because of the Coulomb interaction, the diproton phenomena is extremely hard to observe; the two protons are repelled from one another as soon as they exit the nucleus, making it difficult to observe angular correlations between them.  Nevertheless, it has been observed in many nuclei.  The dineutron decay, on the other hand, poses it own challenges. The neutron dripline is harder to reach than the proton dripline, and the statistics for neutron-rich nuclear decays beyond the neutron dripline, involving two-neutron coincidence, are very low. 
In both, dineutron and diproton decay, the differentiation between a correlated decay and the uncorrelated three-body decay is made based on model considerations and therefore is not free from ambiguity.  

Two-proton decay from the ground state was experimentally observed for the first time in $^{45}$Fe, over forty years after the initial prediction \cite{45FePRL,45FeEur}.  Since then, many examples of two-proton decay have been seen from the ground state \cite{54Zn2p,45Fe48Ni2p,19Mg2p}, as well as from excited states \cite{18Ne2p}.  Because the relevant degrees of freedom are those related to the decay of the two protons, three-body models have been used to theoretically describe these decays.  Different structural configurations of the parent nucleus give rise to different values for the width and half-life, as well as different ways of sharing the energy between the three particles.  It is only through the comparison of model calculations to the data that insights into the nature of the decay can be obtained \cite{Grigorenko2001,Grig2pII}.

In comparison to the large number of two-proton emitters that have been studied experimentally and theoretically, two-neutron emitters have not been as well investigated.  In one of the first theoretical studies of two-neutron decay, Grigorenko \cite{grig20112n} discussed the existence of one-, two-, and four-neutron emitters, as well as comparisons of their widths in a three-body framework.  Recently, a few cases of two-neutron decay have been observed  \cite{Spyrou2012,Kohley2013,Jones2015}. 
The first of these was observed in a 2012 experiment at the National Superconducting Cyclotron Laboratory \cite{Spyrou2012} through the decay of $^{16}$Be to $^{14}$Be plus two neutrons.  As the ground state energy of $^{16}$Be was found to be 1.35 MeV (with a width of 0.8 MeV) and a lower limit of 1.54 MeV had previously been placed on the ground state of $^{15}$Be \cite{Spyrou2011}, $^{16}$Be is an ideal candidate for simultaneous two-neutron decay.  Depending on the width of the ground state of $^{15}$Be, sequential neutron decay from $^{16}$Be to $^{14}$Be could be energetically inaccessible.  A later experiment \cite{Snyder2013} determined that the lowest state in $^{15}$Be is an $l=2$ state at 1.8 MeV with a width of 575 $\pm$ 200 keV.

Although comparisons of the $^{16}$Be data in \cite{Spyrou2012} to dineutron, sequential, and three-body decay models showed the data best matched the dineutron decay, there was some controversy over this finding \cite{Spyrou2012Comm,Spyrou2012Resp}.  Extreme models were used to show the difference between a dineutron decay and a three-body decay.  The dineutron was modeled as a cluster and the decay as $^{16}$Be $\rightarrow$ $^{14}$Be + 2n, in an s-wave relative motion.  The three-body breakup corresponded to phase space only.  A more realistic, full three-body model ($^{14}$Be + n + n) is necessary to help clarify the mode of decay of this exotic nucleus.
Several three-body models have been successfully used to describe the continuum states of $^{26}$O \cite{Grig20132n,Hagino20142n} but no application to $^{16}$Be is thus far available. This is the goal of the present study.

This paper is organized into the following sections.  In Section \ref{theory}, we  introduce the three-body hyperspherical R-matrix theory used in this work.  In Section \ref{numerical}, details about the two- and three-body potentials are presented, as well as a convergence study of our calculations.  Our results, assuming either a $1d_{5/2}$ or a $2s_{1/2}$ ground state for $^{15}$Be, are discussed in Section \ref{results}, and in Section \ref{discussion}, we discuss the consequences of these models.  Finally, we  conclude in Section \ref{conclusions}.


\section{Theoretical framework}
\label{theory}

\begin{figure}[h]
\begin{center}
\includegraphics[width=0.5\textwidth]{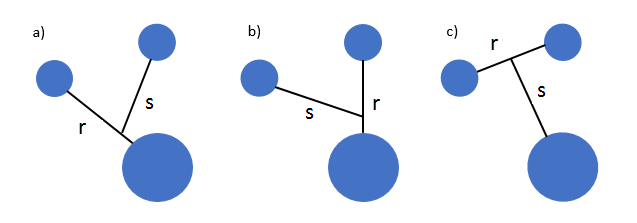}
\end{center}
\caption{Three Jacobi coordinate systems, a) Jacobi X system, b) Jacobi Y system, and c) Jacobi T system.  Because the two neutrons are identical, the X and Y coordinate systems are  identical.}
\label{fig:jacobicoord}
\end{figure}

In this work, the $^{16}$Be system is assumed to take the form of $core+n+n$ and therefore should satisfy the three-body Schrodinger equation:
\begin{equation}
(T_r + T_s + V_{cn_1} + V_{cn_2} + V_{nn} + V_{3b}) \Psi = E_{3B}\Psi \;,
\label{eqn:3b}
\end{equation}
where $\vec r$ an $\vec s$ are the standard Jacobi coordinates, as shown in Figure \ref{fig:jacobicoord}, where $r$ is the distance between two of the particles, and $s$ is the distance between the third particle and the center of mass of the first two.  $V_{cn_i}$ and $V_{nn}$ are the pairwise interactions. 
Typically, when the degrees of freedom in the core are frozen, the final three-body system becomes under-bound.  Traditionally, three-body interactions are then introduced to take into account the additional binding needed to reproduce the experimental ground state. This is the role of $V_{3b}$ in Eqn. (\ref{eqn:3b}).

Eqn. (\ref{eqn:3b}) is a  6-dimensional equation, where the coordinates $\vec r$ and $\vec s$ do not separate due to the fact that the pairwise interactions depend on both. The hyperspherical harmonic method makes a particular choice of coordinates and basis functions such that this three-body Sch\"odinger equation becomes a set of 1-dimensional coupled hyper-radial equations. This is briefly described here.

\subsection{Hyperspherical harmonic method}


For a three-body system, there are three sets of Jacobi coordinates that can be defined, Figure \ref{fig:jacobicoord}. We will use $i$ to denote one of the three Jacobi systems, X, Y, or T. Now, $\vec x$ and $\vec y$  are the scaled Jacobi coordinates \cite{ReactionsBook} (Ch. 9), defined by
\begin{equation}
\vec{x} = \frac{\vec{r}}{\sqrt{2}}
\end{equation}

\noindent and
\begin{equation}
\vec{y} = \sqrt{\frac{2A_3}{A_3+2}}\vec{s},
\end{equation}

\noindent where $A_3$ is the mass number of the core.   From here, we can define the hyperspherical coordinates
\begin{equation}
\rho ^2 = x_i ^2 + y_i ^2,
\label{eqn:rho}
\end{equation}

\noindent and 
\begin{equation}
\mathrm{tan}\theta _i = \frac{x_i}{y_i}.
\label{eqn:theta}
\end{equation}

\noindent Note that $\rho$ is invariant among the three Jacobi coordinate systems, but $\theta$ depends on $i$.  
Using these coordinates, the kinetic energy operator can be written as:
\begin{equation}
\begin{aligned}
T = & -\frac{\hbar ^2}{2m} \left [ \frac{1}{\rho ^5} \frac{\partial}{\partial \rho} \left ( \rho ^5 \frac{\partial}{\partial \rho} \right ) + \frac{1}{\rho ^2 \mathrm{sin}^2 2\theta_i} \frac{\partial}{\partial \theta_i} \left ( \mathrm{sin}^2 2\theta_i \frac{\partial}{\partial \theta_i} \right ) \right. \\
& \left. - \frac{L_x^2}{\rho ^2 \mathrm{sin}^2 \theta_i} - \frac{L_y^2}{\rho ^2 \mathrm{cos}^2 \theta_i} \right ],
\label{eqn:kinop}
\end{aligned}
\end{equation}

\noindent where $m$ is the unit mass, here $m=938.0$ MeV$/\mathrm{c}^2$.

Assuming the $T$ coordinate system for convenience ($i=3$, which we omit for convenience through the rest of this work), we perform the standard partial wave decomposition of the wavefunction,
\begin{eqnarray}
\begin{aligned}
\Psi ^{JM} = & \sum \limits _{l_x l_y lSjI} \psi ^{lSjIJ} _{l_x l_y} (x,y) \\
 & \left \{ \left ( \left [ Y_{l_x} \otimes Y_{l_y} \right ] _l  \otimes \left [ X_{\sigma _1} \otimes X _{\sigma _2} \right ] _S \right ) _j  \otimes \phi _I \right \}_{JM},
\label{eqn:psitot}
\end{aligned}
\end{eqnarray}
\noindent where $l$  is the total orbital angular momentum, $l_x$ is the relative orbital angular momentum in the $2n$ system, $l_y$ is the relative orbital angular momentum in the $core + (2n)$ system, $I$ is the spin of the core, $S$ is the total spin of the two neutrons, and $j$ is the total angular momentum of the two neutrons relative to the core.  
Next we expand the part dependent on $(x,y)$ in hyperspherical functions $\varphi^{l_x l_y}_K(\theta)$,
\begin{equation}
\psi ^{lSjIJ} _{l_x l_y} (x,y) = \rho ^{-5/2} \sum \limits _K ^{K_{max}}\chi ^{lSIjJ} _{K l_x l_y} (\rho) \varphi ^{l_x l_y} _{K} (\theta)
\label{eqn:radpsi}
\end{equation}

\noindent where $\varphi ^{l_x l_y} _{K} (\theta)$ is set to an eigenfunction of the angular operator in Eqn. (\ref{eqn:kinop}) with eigenvalue $K(K+4)$. Its explicit form is:
\begin{eqnarray}
\varphi ^{l_x l_y} _{K} (\theta) = N^{l_x l_y}_{K} (\mathrm{sin}^2 \theta)^{l_x} (\mathrm{cos}^2\theta)^{l_y} P_n ^{l_x + 1/2,l_y+1/2}(\mathrm{cos}2\theta),
\end{eqnarray}
\noindent where $P_n ^{l_x + 1/2,l_y+1/2}(\mathrm{cos}2\theta)$ are the Jacobi Polynomials and $N^{l_x l_y}_{K}$ is a normalization factor resulting from the condition:
\begin{eqnarray}
\int \limits _0 ^{\pi/2} \varphi ^{l_x l_y} _{K} (\theta) \varphi ^{l_x l_y} _{K ^\prime} (\theta) ( \mathrm{sin}^2\theta)  ^{l_x} ( \mathrm{cos}^2 \theta) ^{l_y} d\theta = \delta _{KK^\prime}.
\end{eqnarray}  

For compactness, we introduce the hyperspherical harmonic functions,
\begin{eqnarray}
\begin{aligned}
\mathcal{Y}_\gamma (\Omega _5, \sigma _ 1,\sigma_2,\boldsymbol{\xi}) & = \varphi ^{l_x l_y} _{K} (\theta)\\
& \left \{ \left ( \left [ Y_{l_x} \otimes Y_{l_y} \right ] _l \otimes \left [ X_{\sigma _1} \otimes X _{\sigma _2} \right ] _S \right ) _j \otimes \phi _I \right \}_{JM},
\end{aligned}
\end{eqnarray}

\noindent with $\gamma$ representing the set $\{KlSIjl_xl_y\}$, so that the total wave function can be written in the form:
\begin{eqnarray}
\Psi ^{JM} = \rho ^{-5/2} \sum \limits \chi _\gamma (\rho)\mathcal{Y}_\gamma (\Omega _5, \sigma _ 1,\sigma_2,\boldsymbol{\xi})\;.
\label{eq:wfn}
\end{eqnarray}

\noindent In this work, we focus on $(J,M)=(0,0)$, corresponding to the spin of the ground state.

Substituting Eqn.(\ref{eq:wfn})  into Eqn. (\ref{eqn:3b}), we are left with the following  set of coupled hyper-radial equations:
\begin{eqnarray}
\begin{aligned}
\left ( -\frac{\hbar ^2}{2m} \left [ \frac{d^2}{d\rho ^2} - \frac{(K+3/2)(K+5/2)}{\rho^2} \right ] - E_{3B} \right ) \\
 + \sum \limits _{\gamma ^\prime} V_{\gamma \gamma ^\prime} (\rho) \chi ^J _{\gamma ^\prime}(\rho) = 0,
\label{eqn:radSE}
\end{aligned}
\end{eqnarray}

\noindent where the coupling potentials are defined as 
\begin{eqnarray}
V_{\gamma \gamma ^\prime}(\rho) =
\langle \mathcal{Y}_{\gamma^\prime} (\Omega _5, \sigma _ 1,\sigma_2,\boldsymbol{\xi})|  \sum \limits _{j > i = 1}^3 V_{ij} | \mathcal{Y}_{\gamma} (\Omega _5, \sigma _ 1,\sigma_2,\boldsymbol{\xi}) \rangle.
\end{eqnarray}  

Eqn. (\ref{eqn:radSE}) must be solved under the condition that the wavefunction be regular at the origin and behaves asymptotically as,
\begin{eqnarray}
\chi ^L _{\gamma \gamma _i} \rightarrow \frac{i}{2} \left [ \delta _{\gamma \gamma _i} H^- _{K+3/2}(\kappa \rho) - \textbf{S}^L _{\gamma \gamma _i} H^+ _{K+3/2} (\kappa \rho) \right ],
\label{eqn:scatbc}
\end{eqnarray}
when $\rho \rightarrow \infty$, where the $\gamma_i$ are the components of a plane wave.

It is important also to note that the final wave function will have to be summed over $\gamma_i$, as we do not assume a specific incoming wave for our $^{16}$Be system.

\subsection{Hyperspherical R-matrix method}

The set of coupled hyper-radial equations could, in principle, be solved by direct numerical integration.  However, at low scattering energies, the centrifugal barrier - $(K+3/2)(K+5/2)$ - found in every channel, including $K=0$, would likely cause this method to develop numerical inaccuracies.  Instead, we use the hyperspherical R-matrix method \cite{ReactionsBook} (Ch. 6).   

In the hyperspherical R-matrix method, we first create a basis, $w_{\gamma}^n$, by solving the uncoupled equations, corresponding to Eqn. (\ref{eqn:radSE}) with all couplings set to zero except for the diagonal, in a box of size $\rho_{max}$,
\begin{eqnarray}
\left [ T_{\gamma L}(\rho) + V_{\gamma \gamma} (\rho) - \varepsilon _{n\gamma} \right ] w_{\gamma}^n (\rho) = 0.
\label{eqn:weqn}
\end{eqnarray}

\noindent By enforcing all logarithmic derivatives,
\begin{eqnarray}
\beta = \frac{d \mathrm{ln} (w^n_{\gamma} (\rho))}{d\rho},
\label{eqn:beta}
\end{eqnarray}
\noindent to be equal for $\rho=\rho _{max}$, the set of functions, $w_{\gamma}^n$, form a complete, orthogonal basis within the box.
Then, the scattering equation inside the box can be solved by expanding in this R-matrix basis:
\begin{eqnarray}
g^p _{\gamma} (\rho) = \sum \limits _{n=1} ^N c_{\gamma} ^{pn} w_{\gamma} ^n (\rho).
\label{eqn:gbasis}
\end{eqnarray}

The corresponding coupled channel equations are:
\begin{eqnarray}
\begin{aligned}
\left [ T_{\gamma L}(\rho) + V_{\gamma \gamma} (\rho) \right ] g_{\gamma}^p (\rho)& &  \\
+ \sum \limits _{\gamma ^\prime \ne \gamma} V_{\gamma \gamma ^\prime} (\rho) g^p _{\gamma ^\prime} (\rho)& = & e_p g^p _\gamma (\rho).
\label{eqn:geqn}
\end{aligned}
\end{eqnarray}

To find the coefficients $c_{\gamma}^{np}$, we insert Eqn. (\ref{eqn:gbasis}) into Eqn. (\ref{eqn:geqn}),  multiply the resulting equation by $w_{\gamma ^\prime} ^{n^\prime}$ and integrate over the box size.  This results in a matrix equation:
\begin{eqnarray}
\begin{aligned}
\varepsilon _{n\gamma} c_{\gamma} ^{np} + \sum \limits _{\gamma ^\prime \ne \gamma} \sum \limits _{n^\prime} \langle w_\gamma ^n (\rho) | V_{\gamma \gamma ^\prime} (\rho) | w_{\gamma ^\prime} ^{n^\prime} (\rho) \rangle \\
= e_p c_\gamma ^{pn},
\end{aligned}
\end{eqnarray}
\noindent  which, when solved, provides the coefficients $c_{\gamma} ^{np}$ of the expansion Eqn. (\ref{eqn:gbasis}). Since $g^p _\gamma (\rho)$ are only complete inside the box, and do not have the correct normalization, the full three-body scattering wavefunction is given by a superposition of these solutions which is then matched to the correct asymptotic form:
\begin{eqnarray}
\chi _{\gamma \gamma _i} (\rho) = \sum \limits _{p=1} ^P A_{\gamma _i} ^p g_{\gamma}^p (\rho)\;.
\label{eqn:scatsln}
\end{eqnarray}
\noindent The new expansion parameter $p$ corresponds to the number of poles considered in the R-matrix.
The normalization coefficients, $A_{\gamma _i} ^p$, connect the inside wavefunction with the asymptotic behavior of Eqn. (\ref{eqn:scatbc}). The explicit relation is \cite{ReactionsBook} (Ch. 6),
\begin{eqnarray}
\begin{aligned}
A_{\gamma _i} ^p = \frac{\hbar ^2}{2 \mu} \frac{1}{e_p - E} \sum \limits _{\gamma ^\prime} g^p _{\gamma ^\prime} (\rho_{max} ) \nonumber \\
 \left [ \delta _{\gamma \gamma ^\prime} \left ( H ^{-\prime}_L (\kappa _{\gamma ^\prime}\rho_{max} ) - \beta H^- _L (\kappa _{\gamma ^\prime}\rho_{max} ) \right ) \right. \\
 \left. - \textbf{S}_{\gamma ^\prime \gamma _i} \left ( H ^{+\prime}_L (\kappa _{\gamma ^\prime}\rho_{max}) - \beta H^+ _L (\kappa _{\gamma ^\prime}\rho_{max} ) \right ) \right ].
\label{eqn:a}
\end{aligned}
\end{eqnarray}

From the values of the $g^p_{\gamma}(\rho)$ function at the surface, one can determine the R-matrix \cite{ReactionsBook}, Ch. 6,
\begin{eqnarray}
R_{\gamma \gamma ^\prime} = \frac{\hbar ^2}{2\mu \rho_{max} } \sum \limits _{p=1}^P \frac{g_\gamma ^p(\rho_{max} ) g_{\gamma^\prime} ^p (\rho_{max} )}{e_p-E_{3B}}.
\label{eqn:rmat}
\end{eqnarray}
Once the R-matrix is obtained, the S-matrix can be directly computed:
\begin{eqnarray}
\begin{aligned}
\textbf{S} = \left [ \textbf{H}^+ - \rho_{max} \textbf{R}(\textbf{H}^{+\prime} - \beta \textbf{H}^+) \right ] ^{-1} \\
* \left [  \textbf{H}^- - \rho_{max} \textbf{R}(\textbf{H}^{-\prime} - \beta \textbf{H}^-) \right ] 
\end{aligned}
\end{eqnarray}
along with the phase shifts for each channel, from the diagonal elements of the S-matrix, $S_{\gamma \gamma} = e^{2i\delta _{\gamma \gamma}}$ (more details in \cite{ReactionsBook}).

\subsection{Width calculation}

If one assumes a Breit-Wigner shape, resonant properties for a single-channel calculation can be directly extracted from the phase shift through the relation:
\begin{eqnarray}
\mathrm{tan}\delta = \frac{\Gamma/2}{E_{3B}-E_{res}},
\label{eqn:ps}
\end{eqnarray}
where $\Gamma$ is the width of the resonance and $E_{res}$ is the resonance energy.  If this is valid, the width can be computed as the full width at half maximum (FWHM) from the energy derivative of the phase shift, $\Gamma = \partial \delta /\partial E_{3B}$.  In the case of multiple channels with weak coupling, one can add the various partial widths to obtain the total width of the three-body resonance.
For the strongly coupled three-body problem at hand, we do not expect the pure Breit-Wigner approach to be valid. Nevertheless, for completeness, we do try to identify channels for which such an approach may be applicable. 

We can also construct the total three-body elastic scattering as a function of energy,
\begin{eqnarray}
\frac{d\sigma}{dE_{3B}} = \frac{1}{4\kappa^5} \sum \limits _\gamma \left | 1 - S_{\gamma \gamma} (E_{3B}) \right |^2,
\end{eqnarray}
from which we can extract a resonance energy and width.  Theoretically, three-body elastic scattering could be measured if the $^{14}$Be and two neutrons could be impinged upon one another simultaneously.  This method can justify resonance energies extracted from a single phase shift, as well as lend itself to a width calculation that includes all of the channels.

\section{Numerical Details}
\label{numerical}

\subsection{Input interactions versus data}
\label{input}

\begin{figure}[h]
\begin{center}
\includegraphics[width=0.5\textwidth]{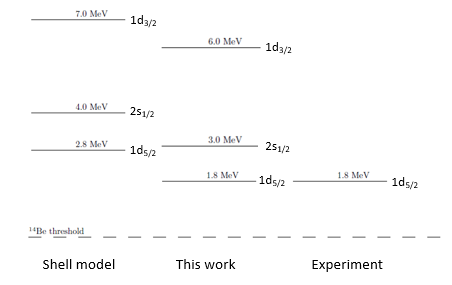}
\end{center}
\caption{Level scheme for $^{15}$Be.  The first column shows the shell model calculation provided by \cite{Browncomm}, while the second column shows the $^{15}$Be levels that we used in this work; here, the shell model levels are lower by 1 MeV so the $1d_{5/2}$ state in the shell model calculation reproduces the experimental $l=2$ energy from \cite{Snyder2013}, as shown in the third column.}
\label{fig:2blevels}
\end{figure}

In the three-body model, each of the two-body interactions must be constrained, typically from experimental data.  However, very little is known about $^{15}$Be \cite{Snyder2013}, so shell model calculations are used to supplement the available data.  Shell model calculations for $^{15}$Be were provided \cite{Browncomm} using the WBP interaction \cite{BrownWBP}.  Since the ground state in the shell model calculation was an $l=2$ state and was 1 MeV higher than the experimentally observed $l=2$ state in $^{15}$Be \cite{Snyder2013}, the levels that were used to constrain the $^{14}$Be-n interactions were the shell models levels lowered by 1 MeV, shown in Figure \ref{fig:2blevels}.

The $^{14}$Be-n interaction for each partial wave has a Woods-Saxon shape with $a=0.65$ fm and $R=1.2A^{1/3}$ fm, where $A$ is the mass number of the $^{14}$Be core.  The depths depend on angular momentum, and are obtained by fitting the single-particle resonances in $^{15}$Be, described in Fig. \ref{fig:2blevels}, using the code {\sc poler} \cite{poler}.  The core deformation is taken into account by allowing an $l$-dependence in the potential.  A spin-orbit interaction was also included with the same geometry as the central nuclear force with the depth adjusted to reproduce the split between the $1d_{5/2}$ and $1d_{3/2}$ states. We use the definition of the spin-orbit strengths of FaCE \cite{FaCE}.  Potential depths for the various models included are as indicated in Table \ref{tab:d52parm}.

The lowest s- and p-orbitals in $^{14}$Be are assumed to be full. In order to remove the effect of these occupied states in the $^{14}$Be core, the $1s_{1/2}$, $1p_{3/2}$, and $1p_{1/2}$ states were projected out through a supersymmetric transformation \cite{FaCE}.  

\subsection{Description of models}

\begin{table}[b]
\centering
\begin{tabular}{| c | c | c | c | c | c |}
\hline \textbf{Parameter} & \textbf{D3B} & \textbf{D} & \textbf{DNN}  & \textbf{S}\\ \hline
$V_s$ & --26.182 & --26.182 & --26.182 & --41.182 \\ \hline
$V_p$ & --30.500 & --30.500 & --30.500 & 30.500 \\ \hline
$V_d$ & --42.73 & --42.730 & --42.730 &  --42.730 \\ \hline
$V_{so}$ (l$\ne$2) & --10.000 & --10.000 & --10.00 & --10.000 \\ \hline
$V_{so}$ (l=2) & --33.770 & --33.770 & --33.770 & --33.770 \\ \hline
$V_{3B}$ & --7.190  & 0.000 & --7.190 & 0.000 \\ \hline
$\alpha_{NN}$ & 1.000 & 1.000 & 0.000 & 1.000 \\ \hline
\end{tabular}
\caption{Interaction parameters for the various models considered. All depths are given in MeV.  Details in the text.}
\label{tab:d52parm}
\end{table}

There are four three-body models for $^{16}$Be that we consider in this work. 
In {\bf D3B}, the ground state of $^{15}$Be is a $1d_{5/2}$ state and a three-body force is included to reproduce the experimental three-body ground state energy of $^{16}$Be.   This three-body force is also of Woods-Saxon form with radius of $3.02$ fm and diffuseness of $0.65$ fm.  In {\bf D}, the ground state of $^{15}$Be is a $1d_{5/2}$ state but no three-body force is included.  In {\bf S}, the ground state of $^{15}$Be is a $2s_{1/2}$ state but no three-body force is included.  

All models D3B, D, and S include the GPT NN interaction \cite{gpt}, as in previous three-body studies \cite{Nunes1996,Timo2ncore,Brida2008,Brida2006}.  This interaction reproduces NN observables up to 300 MeV.  Although it is simpler than the AV18 \cite{AV18} and Reid soft-core \cite{Reid} interactions, its range is more than suitable for the energy scales used in this work.  
We also consider the effects of removing the NN interaction completely. This model is named \textbf{DNN}. In Table \ref{tab:d52parm} we provide the depths for the various terms of the interaction and the coefficient $\alpha_{NN}$ by which we multiply the GPT force in each of our calculations.

\begin{table}[h]
\centering
\begin{tabular}{| c | c | c | c | c |}
\hline& \textbf{D3B} & \textbf{D} & \textbf{DNN}  & \textbf{S}\\ \hline
$^{15}Be(1d_{5/2})$ & 1.80 & 1.80 & 1.80 & 1.80 \\ \hline
$^{15}Be(2s_{1/2})$ & $\sim$3 & $\sim$3 & $\sim$3 & 0.48 \\ \hline
$^{16}Be(gs) $ & 1.35 & 1.84 & 3.14 & ---  \\ \hline
\end{tabular}
\caption{Energy levels, in MeV, for $^{16}$Be and $^{15}$Be for the various models considered. Energies are measured with respect to the $^{14}$Be threshold.  Details in the text.}
\label{tab:energies}
\end{table}

In Table \ref{tab:energies}, we summarize the energies for the $1d_{5/2}$ and $2s_{1/2}$ states in the subsystem $^{15}$Be as well as the ground state energy of $^{16}$Be in the various models considered in Table \ref{tab:d52parm}.  For all of the models considered, the $1d_{3/2}$ state was placed at 6.0 MeV.

\subsection{Convergence}

\begin{figure}[h]
\begin{center}
\includegraphics[width=0.5\textwidth]{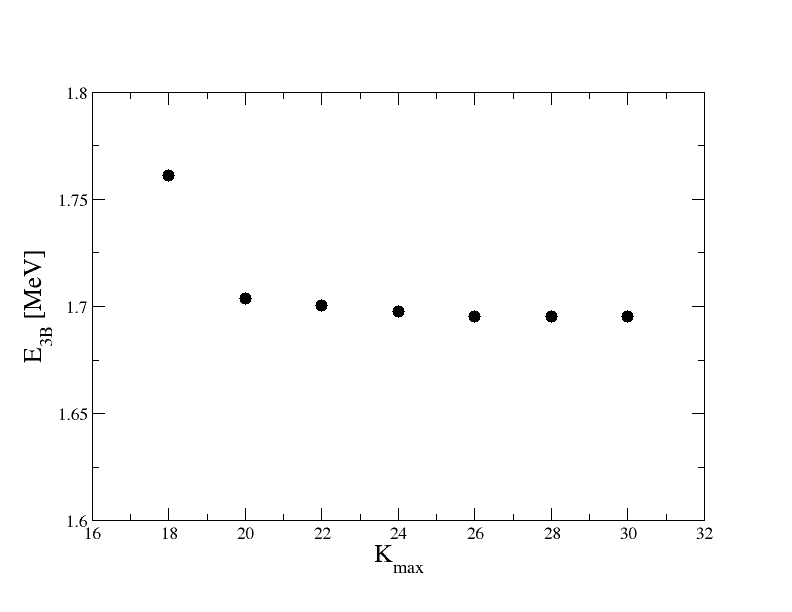}
\end{center}
\caption{Convergence of the three-body energy as a function of the maximum K value included in the model space.}
\label{fig:convg_EKmax}
\end{figure}

Our methods rely on basis expansions, and our model space is determined by a number of numerical parameters. In this section we demonstrate convergence for various quantities, including the ground state energy of $^{16}$Be  and the phase shifts.  The truncation of the  expansion in hyperspherical harmonics is controlled by the hyper-momentum, $K$. In  Figure \ref{fig:convg_EKmax}, we show the convergence of the lowest $0^+$ three-body resonance energy of $^{16}$Be as $K_{max}$  increases.  The width of $^{16}$Be with respect to $K_{max}$ shows the same trend.  Our results are converged within $0.05$ MeV by $K_{max}=28$ for both observables.
 
Our results are very sensitive to the the number of R-matrix basis functions $N$ (which essentially determines the hyper-radial discretization) as well as the maximum box size, $\rho_{max}$. In Tables \ref{tab:convg_nsturm} and \ref{tab:convg_rmax}, we show the convergence of the three-body resonance energy of several parameters for the $K=0, L=0, S=0$ channel.  The convergence with respect to the number of R-matrix basis functions is shown in Table \ref{tab:convg_nsturm}. Convergence is slow but results are very close to converged for $N=95$.

We also needed to check the dependence on the box size, $\rho_{max}$. When increasing the box size, one also needs to increase the number of R-matrix basis functions that span the radial space for consistency. These results are given in Table \ref{tab:convg_rmax}.  We summarize the minimum convergence requirements in Table \ref{tab:d52convg}.

\begin{table}[h]
\centering
\begin{tabular}{| c | c |}
\hline \textbf{N} & \textbf{E$_{3B}$ (MeV)} \\ \hline
70 & 2.06 \\ \hline
75 & 1.95 \\ \hline
80 & 1.84 \\ \hline
85 & 1.78 \\ \hline
90 & 1.74 \\ \hline
95 & 1.71 \\ \hline
100 & 1.69 \\ \hline
105 & 1.67 \\ \hline
\end{tabular}
\caption{Convergence of E$_{3B}$ as a function of the number of radial R-matrix functions, N, for $\rho_{max}=60$ fm.}
\label{tab:convg_nsturm}
\end{table}

\begin{table}[h]
\centering
\begin{tabular}{| c | c | c |}
\hline \textbf{$\rho_{max}$ (fm)} & \textbf{N} & \textbf{E$_{3B}$ (MeV)} \\ \hline
50 & 80 & 1.70 \\ \hline
60 & 95 & 1.71 \\ \hline
70 & 110 & 1.72 \\ \hline
\end{tabular}
\caption{As the box size $\rho_{max}$ increases, a greater number of R-matrix radial functions, N, are need to keep the same resonance energy, E$_{3B}$.}
\label{tab:convg_rmax}
\end{table}

\begin{table}[h]
\centering
\begin{tabular}{| c | c |}
\hline \textbf{Parameter} & \textbf{Value} \\ \hline
K$_{\mathrm{max}}$ & 28 \\ \hline
l$_x(max)$, l$_y(max)$ & 10 \\ \hline
$N_{Jac}$ & 65 \\ \hline
$\rho_{max}$ (fm) & 60 \\ \hline
N & 95 \\ \hline
\end{tabular}
\caption{Minimum convergence values for the three-body wave function expansion.}
\label{tab:d52convg}
\end{table}


\section{Results}
\label{results}

\begin{figure}[h]
\begin{center}
\includegraphics[width=0.5\textwidth]{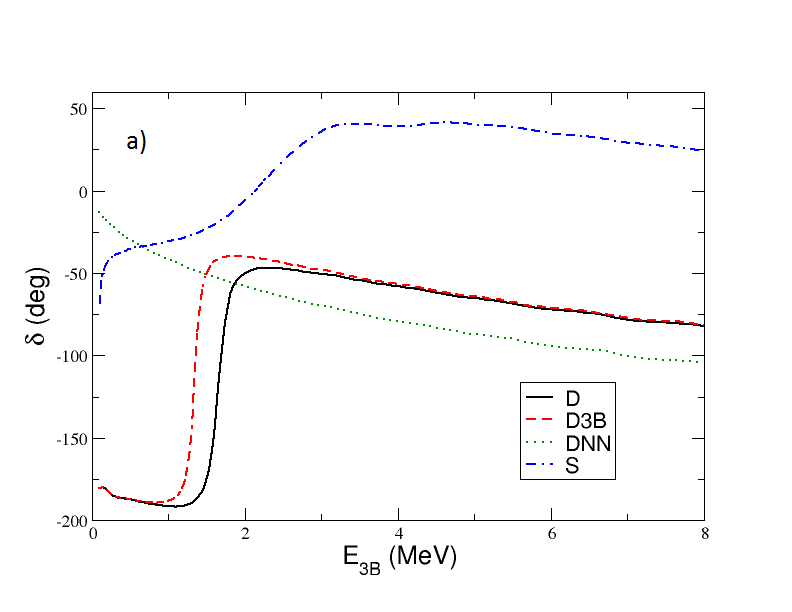}
\includegraphics[width=0.5\textwidth]{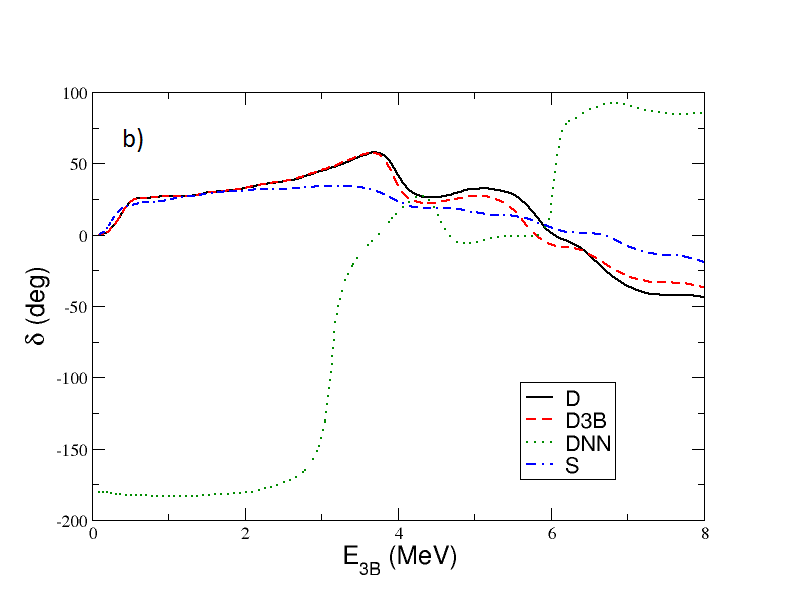}
\includegraphics[width=0.5\textwidth]{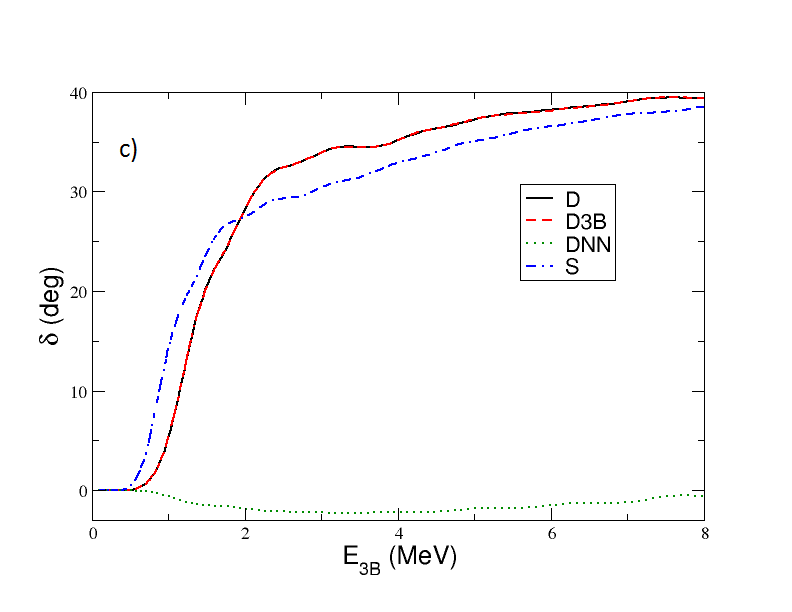}
\end{center}
\caption{Phase shifts as a function of three-body energy for $^{16}$Be for models D (solid, black), D3B (dashed, red), DNN (dotted, green), S (double-dash dotted, blue).  Panel (a) refers to $K=0$, $l_x=0$, $l_y=0$ channel;  panel (b) refers to $K=4$, $l_x=0$, $l_y=0$ channel and panel (c) refers to $K=10$, $l_x=0$, $l_y=0$ channel; }
\label{fig:phase}
\end{figure}

Using model D, we calculated the phase shifts for $^{16}$Be.  The converged phase shift as a function of the three-body energy for the $K=0$ channel is found in Figure \ref{fig:phase} panel (a), solid line.  As we would expect for this type of system, the resonance energy in model D is above the experimental energy observed for the ground state.  We include a three-body force, as described in Table \ref{tab:d52parm}.  The phase shift for the $K=0$ channel, including this three-body interaction (model D3B), is shown in Figure \ref{fig:phase}, panel (a), dashed line.

Figure \ref{fig:phase} shows not only the component with the lowest hypermomentum but also a few other components for illustration purposes:  (a) $K=0, l_x=l_y=0$, (b) $K=4, l_x=l_y=0$, and  (c) $K=10, l_x=l_y=0$.  While the $K=0$ channel contains a very clear signal of the resonance, other channels also contribute. This can be verified by the behavior of the phase shift around the resonance energy in the $K=4$ and $K=10$ channels.  In particular, the $K=10$, $l_x=l_y=0$ in Figure \ref{fig:phase} (c) shows a broader contribution to the resonance, which should have a much larger contribution to the width of $^{16}$Be in D3B.  However, the shape of the resonance is not a simple Breit-Wigner as in (a).  Therefore, simply calculating the widths from each of the channels and adding them together to find a total width is not straightforward.  

\begin{figure}[h]
\begin{center}
\includegraphics[width=0.5\textwidth]{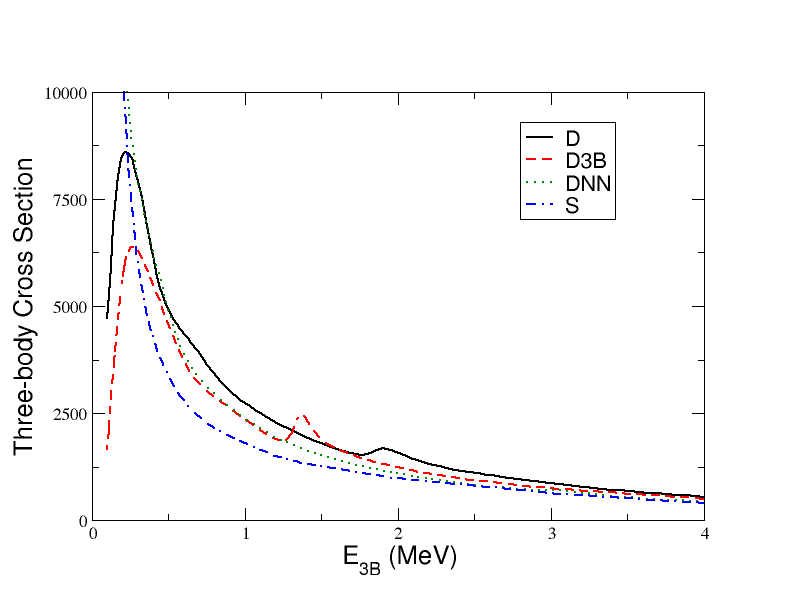}
\end{center}
\caption{Three-body cross section as a function of three-body energy for $^{16}$Be models D(solid, black), D3B (dashed, red), DNN (dotted, green), S (double-dash dotted, blue).}
\label{fig:crosssection}
\end{figure}

Instead, one can extract a width from the three-body elastic cross section, shown in Figure \ref{fig:crosssection} as a function of three-body energy.  This observable contains not only the contribution from the $K=0$ channel but all of the other channels included in the model space.  
If one investigates the structure of the wavefunction of model D3B for the pole closest to the resonance energy, we conclude that the state is 37\% $K=0$, $l_x=l_y=0$, 30\% $K=2, l_x=l_y=0$, and 13\% $K=4$, $l_x=l_y=0$.  
 

Although the lowest experimentally observed state in $^{15}$Be was an $l=2$ state, we wanted to investigate the possibility of an s-wave ground state in $^{15}$Be, below the observed state.  Such a state exists in $^{10}$Li and was only observed after other higher lying resonances were well known \cite{Thoennessen1999}. With this in mind, we developed model S,  described in Table \ref{tab:d52parm}.  We use the same model space as in Table \ref{tab:d52convg}.  
The dot-dashed line in Figure \ref{fig:phase} shows the corresponding phase shifts for several components of the wavefunction.  The resulting cross section is also depicted in Figure \ref{fig:crosssection} by the dot-dashed line.  The clear evidence for the resonance seen in models D and D3B, is washed out in model S.  We will come back to this  in Section \ref{discussion}.

Finally we also consider the results when the NN interaction is switched off (model DNN). Then the resonance disappears from the $K=0$ phase shift, and instead appears in the $K=4$ channel at around $3$ MeV. This demonstrates the importance of the NN correlation to produce the observed state in $^{16}$Be.  Our results show that the configuration of the system is strongly modified by switching off the NN interaction.


\section{Discussion}
\label{discussion}

\begin{figure}[h]
\begin{center}
\includegraphics[width=0.5\textwidth]{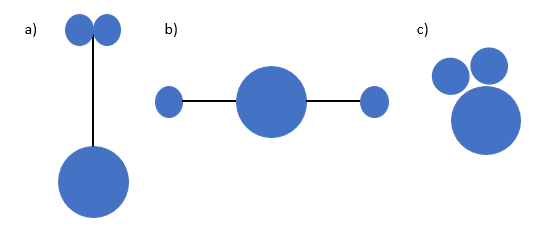}
\end{center}
\caption{Three-body configurations, a) dineutron (two neutrons close together and far from the core), b) helicopter (two neutrons are close to the core and far from each other), and c) three-body (the three bodies are equally spaced).}
\label{fig:3bconfig}
\end{figure}

In calculating the spatial probability distribution of the three-body system,
\begin{equation}
P(x,y) = \int |\Psi ^{JM} (\textbf{x},\textbf{y})|^2 d\Omega _x d\Omega _y
\end{equation}
\noindent we can determine the location of the two neutrons with respect to the core.  
From this density distribution we can determine the configuration of the two neutrons in $^{16}$Be - dineutron, helicopter, or triangle (Figure \ref{fig:3bconfig} a, b, and c, respectively).  

\begin{figure}[h]
\begin{center}
\includegraphics[width=0.5\textwidth]{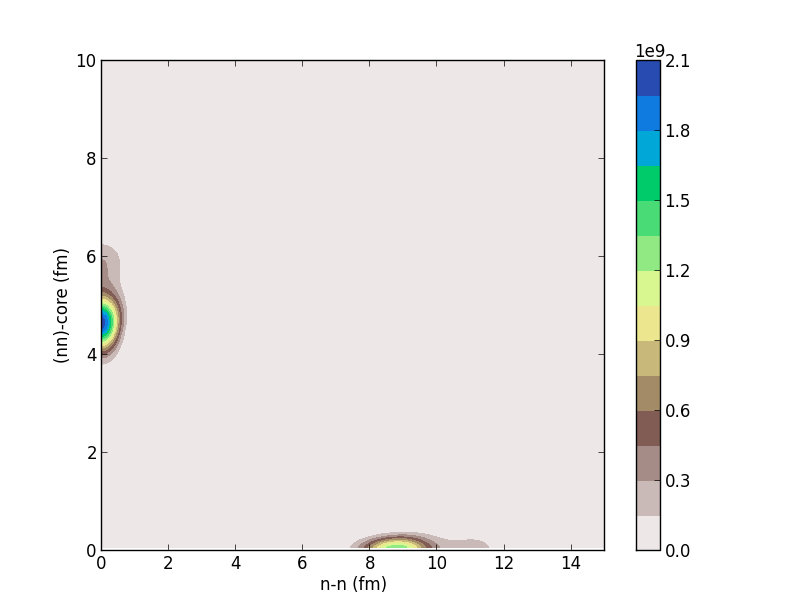}
\end{center}
\caption{Three-body density as a function of the distance between the two neutrons ($r$) and the distance between the nn pair and the core ($s$) for D3B.  The scale on the right is given in fm$^{-5}$.}
\label{fig:rsden_d52}
\end{figure}

Figure \ref{fig:rsden_d52} shows the resulting density distribution for the $^{16}$Be system with the D3B model.  The density distribution mainly shows a dineutron configuration, although a small component of a helicopter configuration is present.  This is consistent with what was seen in \cite{Spyrou2012}.  Even though the three-body resonance energy shifts up by about 0.5 MeV when the three-body interaction is removed, this does not change the relative strength of the dineutron component of the density distribution.

There are several quantities that we can look at to extract a width for this system.  If we extract the width from the FWHM of the derivative of the three-body phase shift in Figure \ref{fig:phase} (a) we obtain $0.16$ MeV (consistent with the width of the nearest R-matrix pole, $0.17$ MeV).  We can also look at the width that would be extracted from the elastic cross section, Figure \ref{fig:crosssection}.  To remove some of the background from the peak, we subtract the elastic cross section for model DNN since they have roughly the same magnitude and shape.  The width from the FWHM of the peak is $0.16$ MeV, consistent with the other two calculations.  However, this is smaller than the $0.8$ MeV width found by experiment \cite{Spyrou2012}.  This discrepancy is most likely due to the effect of experimental resolution, efficiencies, and acceptances of the detector set up, which has not been taken into account in these calculations.  Work to include these effects is currently ongoing.

\begin{figure}[t]
\begin{center}
\includegraphics[width=0.5\textwidth]{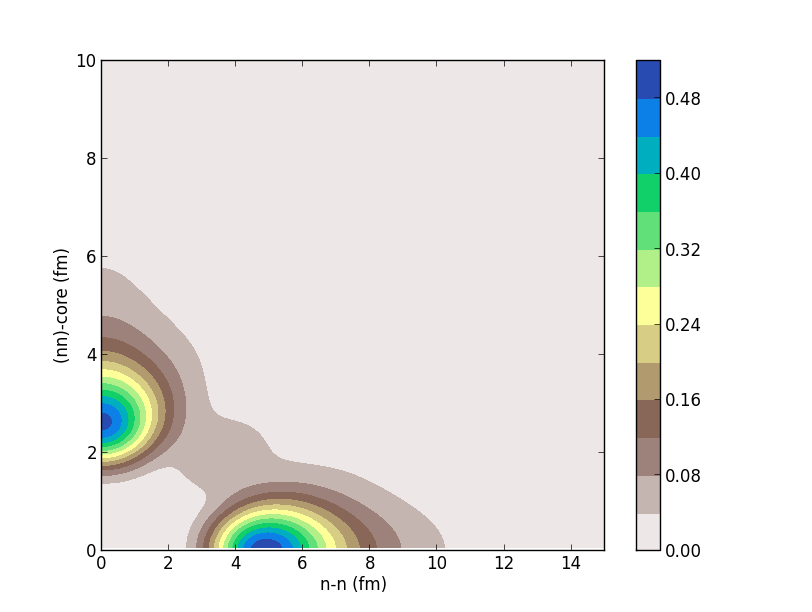}
\end{center}
\caption{Same as Fig. \ref{fig:rsden_d52} for the model DNN.}
\label{fig:rsden_vnn}
\end{figure}

When we switch off the NN interaction (model DNN), the density distribution shown in Figure \ref{fig:rsden_vnn} has equal contributions from the dineutron and the helicopter configurations.  Increasing or decreasing the strength of the three-body interaction does not change this picture.  This illustrates that it is indeed the NN interaction that is responsible for the strong dineutron character of the $^{16}$Be ground state.

\begin{figure}[h]
\begin{center}
\includegraphics[width=0.5\textwidth]{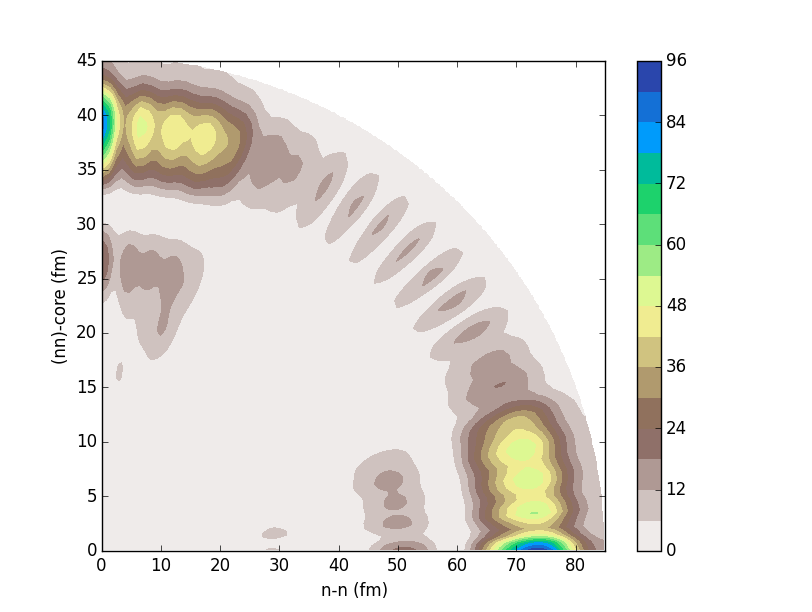}
\end{center}
\caption{Same as Fig. \ref{fig:rsden_d52} for a plane wave solution of $^{16}$Be, for comparison.}
\label{fig:rsden_planewave}
\end{figure}

For comparison with all of these models, Figure \ref{fig:rsden_planewave} shows the density distribution for a $^{16}$Be that has both the NN and n-$^{15}$Be interactions removed.  This system does not contain any resonance, so the density distribution is calculated at an arbitrary energy, 6.347 MeV.  The distribution has less structure and is pushed farther away from the center of the system.

Let us now turn our attention to the hypothesis of there being a lower s-wave resonance in $^{15}$Be (model S).
As shown in Figure \ref{fig:crosssection}, no clear signature of a resonance was found in the elastic cross section. Indeed the $^{16}$Be system becomes bound.
Only by using a much shallower s-wave potential could we regain a resonance in the low energy $^{16}$Be spectrum. These results make it much less likely that $^{15}$Be has an s-wave ground state.

Using three-neutron coincidences, Kuchera, et. al. proposed that there is a small chance of finding the $1d_{3/2}$ state in $^{15}$Be at 2.69 MeV \cite{Kuchera2015}.  Including this state, keeping the $1d_{5/2}$ at 1.8 MeV, and using the same s-wave as models D and D3B, the ground state energy of $^{16}$Be was at $1.06$ MeV, without including a three-body interaction.  The density distribution was nearly identical to that shown in Figure \ref{fig:rsden_d52}. In this case, the only way to reproduce the experimental ground state of $^{16}$Be would be to include a repulsive three-body interaction, which is unusual.


\section{Conclusions}
\label{conclusions}

In summary, a three-body model for  $^{16}$Be was developed to investigate the properties of the system in the continuum.  The hyperspherical R-matrix method was used to solve the three-body scattering problem, with the n-$^{14}$Be interactions constrained by experimental data on $^{15}$Be.  As usual in three-body models, we included a three-body potential to reproduce the experimental ground state energy of $^{16}$Be. We obtained convergence results for phase shifts, density distributions and elastic cross sections. 

We study the properties of the resulting three-body continuum around the resonant energy of $^{16}$Be and conclude that it has a strong dineutron configuration, consistent with experimental observations \cite{Spyrou2012}. The estimate of the width obtained from our calculations is consistent among the various methods of extraction but is smaller than the experimental value \cite{Spyrou2012}.  We find that the NN interaction is important in producing the strong dineutron configuration in the ground state of $^{16}$Be, since the structure of the resonance is completely different when switching off the NN interaction. In contrast, the three-body force needed to shift the resonance energy to the observed experimental energy of $^{16}$Be ground state, has little effect on the structure of the state. We also explore a possible s-wave ground state in $^{15}$Be and find that the results are incompatible with the observed $^{16}$Be ground state \cite{Spyrou2012}. In fact, the structure of the $^{15}$Be ground state being a $d_{5/2}$-wave  is crucial to reproducing the $^{16}$Be experimental results.  Only with $1d_{5/2}$ ground state and higher lying $2s_{1/2}$ and $1d_{3/2}$ state in $^{15}$Be can a resonance energy of 1.35 MeV be reproduced in $^{16}$Be with a physical three-body interaction.

The $^{16}$Be experiment \cite{Spyrou2012} provided a variety of correlation observables that would be very interesting to compare to our model. However, our predictions need to be introduced into a full experimental simulation code that includes the appropriate three-body assumptions as well as efficiencies and acceptance of the detector setup. Work along these lines is currently underway.
 

\begin{center}
\textbf{ACKNOWLEDGMENTS}
\end{center}

The authors would like to thank Artemis Spyrou, Michael Thoennessen, Simin Wang, and Anthony Kuchera for useful discussions.  The authors would also like the acknowledge iCER and the High Performance Computing Center at Michigan State University for their computational resources.  This work was supported by the Stewardship Science Graduate Fellowship program under Grant No. DE-NA0002135 and by the National Science Foundation under Grant. No. PHY-1403906 and PHY-1520929.  This work was performed under the auspices of the U.S. Department of Energy through NNSA contract DE-FG52-08NA28552 and by Lawrence Livermore National Laboratory under Contract DE-AC52-07NA27344.




\bibliography{16Benn}

\end{document}